\documentclass[aps,prd,twocolumn,epsf,nofootinbib]{revtex4}
\usepackage{amsfonts,latexsym,eucal,amsmath,amssymb,times,mathrsfs}
\usepackage[dvips]{graphicx}
\usepackage{color}

\newcommand{\bm}[1]{\hbox{\boldmath{$#1$}}}

\newcommand{\bk}{\mbox{\boldmath{\scriptsize $k$}}}
\newcommand{\bx}{\mbox{\boldmath{\scriptsize $x$}}}
\newcommand{\Mp}{M_{\rm pl}}
\newcommand{\dd}{{\rm d}}
\newcommand{\eps}{\varepsilon}
\newcommand{\sR}{{^s\!R}}
\newcommand{\sbR}{{^s\!\breve{R}}}
\newcommand{\gR}{{^g\!R}}
\newcommand{\Approx}{\stackrel{\tiny\rm IR}{\approx}}

\begin{document}

\thispagestyle{empty}


\title{Natural selection of inflationary vacuum required by
infra-red regularity and gauge-invariance}
\date{\today}
\author{Yuko Urakawa$^{1}$}
\email{yuko_at_gravity.phys.waseda.ac.jp}
\author{Takahiro Tanaka$^{2}$}
\email{tanaka_at_yukawa.kyoto-u.ac.jp}
\address{\,\\ \,\\
$^{1}$ Department of Physics, Waseda University,
Ohkubo 3-4-1, Shinjuku, Tokyo 169-8555, Japan\\
$^{2}$ Yukawa Institute for Theoretical Physics, Kyoto university,
  Kyoto, 606-8502, Japan}

\preprint{2010-**-**, WU-AP/***/, hep-th/*******}


\begin{abstract}
It has been an issue of debate whether the inflationary infrared(IR) divergences are 
physical or not. Our claim is that, 
at least, in single-field models, the answer is 
 ``No,'' and that the spurious IR divergence is originating from the
 careless treatment of the gauge modes. 
In our previous work we have explicitly shown
 that the IR divergence is absent in the genuine gauge-invariant quantity
 at the leading order in the slow-roll approximation. 
We extend our argument to include higher-order slow-roll corrections and the
 contributions from the gravitational waves. The key issue is to assure
 the gauge invariance in the choice of the initial vacuum, which is a
 new concept that has not
 been considered in conventional calculations.  
\end{abstract}


\pacs{98.80.-k, 98.80.Bp, 98.80.Cq, 04.20.Cv}
\maketitle


\section{Introduction}
The importance of the gauge-invariant perturbation has been widely
recognized since decays ago, but some subtle issues get in the way of
its realization particularly in non-linear perturbation theory. 
During inflation, massless fields are known to
yield the scale invariant spectrum $P(k) \propto 1/k^3$ at linear
order. 
These fields contribute to the one-loop diagram with the four point
interaction as $\int \dd^3\!k/k^3$ in the long wavelength limit, which leads to the logarithmic
divergence~
\cite{Boyanovsky:2004gq, Boyanovsky:2004ph, Boyanovsky:2005sh,
Boyanovsky:2005px, Onemli:2002hr, Brunier:2004sb, Prokopec:2007ak,
Sloth:2006az, Sloth:2006nu, Seery:2007we, Seery:2007wf, Urakawa:2008rb,
Adshead:2008gk, Cogollo:2008bi, Rodriguez:2008hy, Seery:2009hs, Gao:2009fx, Bartolo:2010bu, Seery:2010kh, Kahya:2010xh}.
In our previous work~\cite{IRsingle}, we pointed out that,
at least in single field models, the IR divergences are attributed 
to the bad treatment of gauge degrees of freedom. The gauge
degrees of freedom can be classified into two classes: 
the local ones and the non-local ones. In the usual calculation 
only the former is fixed by adapting particular
gauge conditions at each space-time point. However, this is not
sufficient in order to accomplish the complete gauge 
fixing because of the presence of the non-local gauge degrees 
of freedom, which 
are typically the degrees of freedom to specify boundary conditions 
in solving the lapse function and the shift vector. 
These non-local gauge degrees of freedom are formally fixed 
by imposing the regularity at spatial infinity
in the conventional perturbation theory. 
As a result, however, the time evolution of so-called gauge invariant 
variables is affected by the information from infinitely large 
volume outside our observable region. 
We claimed that this is the origin of IR divergences~\cite{IRsingle}.

Along the line mentioned above, 
the IR divergence problem reminds us of the importance of 
maintaining the gauge-invariance in cosmological perturbation. In our previous
work~\cite{IRgauge_L}, we provide one simple but calculable example of genuine
gauge-invariant quantities, and showed its regularity at the leading
order in the slow-roll approximation. 
In order to realize IR regular perturbation theory, one
important additional 
aspect is to guarantee the gauge invariance of the initial
quantum state. We found that by choosing the Bunch-Davies vacuum, which 
yields the scale-invariant spectrum, at the lowest order in 
slow-roll approximation, the gauge invariance of the 
initial quantum state is realized. In this paper, we
extend our argument about IR regularity of such genuine 
gauge invariant quantities  
and the existence of gauge-invariant initial state to
the quadratic order in the slow-roll approximation. 
This extension would be wanted, because the presence of
IR divergences that has been reported so far mostly starts
with this order~(see Ref.~\cite{Seery:2010kh} for a recent review). We also include the discussion 
about the contributions from the graviton loops.

To quantify the primordial fluctuations and provide the testable
predictions for models of inflation, it is necessary to remedy the singular
behaviour of IR corrections as well as the
ultraviolet divergence~\cite{Weinberg:2005vy, Weinberg:2006ac,
Senatore:2009cf}. The feasibility of the secular growth of IR contributions 
has also been addressed, motivated as a possible solution to the
smallness of the cosmological constant~\cite{Tsamis:1996qm,
Tsamis:1996qq, Polyakov:2009nq, Kitamoto:2010si}. (See also
Refs.~\cite{Garriga:2007zk} and \cite{Tsamis:2007is}.) Despite the
several efforts~\cite{IRsingle,IRmulti, Lyth:2007jh,
Bartolo:2007ti, Riotto:2008mv, Enqvist:2008kt, Burgess:2009bs,
Burgess:2010dd, Marolf:2010zp, Rajaraman:2010zx, Rajaraman:2010xd, Giddings:2010nc,
Byrnes:2010yc}, the debate regarding the possibility of the IR
divergence has not been settled. To put an end to this debate, following
the idea presented in
our previous works~\cite{IRsingle, IRgauge_L}, we explicitly
show the absence of the IR divergence, restricting our argument to
single field models of inflation.

Our paper is organized as follows. In Sec.~\ref{Sec:Review}, we give the
setup of our problem and briefly review our solution to the IR
divergence problem. In Sec.~\ref{SSec:originIR}, we clarify 
the relation between the residual gauge degrees of freedom 
and the boundary conditions 
in solving the lapse function and the shift vector. 
In Sec.~\ref{Sec:GIQ}, we give one
example of genuine gauge-invariant variables. 
In Sec.~\ref{Sec:IRregularity}, we show the regularity of
the genuine gauge-invariant variable and study the requirement 
of the gauge-invariance on the initial quantum state. 
Our results are summarized in Sec.~\ref{Sec:Conclusion}.

\section{Brief review of IR divergence problem}
\label{Sec:Review}
In this section, we briefly summarize our solution to the IR divergence
problem, proposed in our previous work~\cite{IRgauge_L}. 

\subsection{Basic equations} 
\label{SSec:Basiceq}
We consider a 
standard single field inflation model whose action takes the form 
\begin{eqnarray}
 S = \frac{\Mp^2}{2} \int \sqrt{-g}~ [R - g^{\mu\nu}\phi_{,\mu} \phi_{,\nu} 
   - 2 V(\phi) ] \dd^4x~, 
\end{eqnarray}
where $\Mp$ is the Planck mass and the scalar field $\phi$ was 
rescaled as $\phi \to \phi/\Mp$ to be dimensionless. The ADM formalism has been utilized
to derive the action of the dynamical variables particularly in the
non-linear perturbation theory~\cite{Maldacena2002}. Using the decomposed metric
\begin{eqnarray}
 \dd s^2 = - N^2 \dd t^2  + h_{ij} (\dd x^i + N^i \dd t) (\dd x^j + N^j
  \dd t)~, 
\end{eqnarray}
the action is rewritten as
\begin{eqnarray}
 S&\!=&\!\frac{\Mp^2}{2} \int\! \sqrt{h} \Bigl[ N \,\sR - 2 N
  V(\phi) + \frac{1}{N} (E_{ij} E^{ij} - E^2) 
 \cr && \qquad \quad
 + \frac{1}{N} ( \partial_t \phi
  - N^i \partial_i \phi )^2 - N h^{ij} \partial_i \phi \partial_j \phi
  \Bigr] \dd^4x~, \nonumber \\
\end{eqnarray}
where $\sR$ is the three-dimensional scalar curvature and $E_{ij}$ and $E$ are defined by 
\begin{eqnarray}
 E_{ij} = \frac{1}{2} \left( \partial_t h_{ij} - D_i N_j
 - D_j N_i \right), \quad E = h^{ij} E_{ij} ~.
\end{eqnarray}
The spatial index of $N_i$ is raised and lowered by $h_{ij}$.

In this paper we work both in the comoving gauge and in 
the flat gauge. 
We defer the introduction of the latter gauge to 
Sec.~\ref{SSec:tranformation}. 
The comoving gauge is defined by 
\begin{equation}
 \delta\phi=0~,
\label{comoving}
\end{equation}
where $\delta\phi$ is the perturbation of the scalar field.   
We decompose the spatial metric as
\begin{eqnarray}
  h_{ij} 
 =  e^{2 (\rho + \zeta) } \left[ e^{\delta \gamma} \right]_{ij}~, 
\label{Cond:comoving} 
\end{eqnarray}
where $a:=e^{\rho}$ denotes the background scale factor, 
and ${\rm tr}[ \delta \gamma ]=0$. 
Using the degrees of freedom in the choice of the spatial coordinates,  
we further impose the gauge conditions 
$\partial^i \delta \gamma_{ij}=0$. 
Here, the indices of spatial derivatives are raised or lowered 
by using Kronecker's delta as $\partial^i=\delta^{ij}\partial_j$.

Varying the action with respect to $N$ and $N^i$, 
we obtain the Hamiltonian and momentum constraints as
\begin{align}
 &  \sR - 2 V -  N^{-2}  (E^{ij} E_{ij} - E^2 )  
 - N^{-2} \left( \partial_t\phi\right)^2  = 0\,, \label{Eq:Hconst} \\
 &  D_j \left[ N^{-1} \left( {E^j}_i - {\delta^j}_i E \right) \right] =
  0\,. \label{Eq:Mconst}
\end{align}
Introducing the perturbed variables as
\begin{eqnarray}
 \check{h}_{ij}:= e^{- 2\rho}h_{ij}, \quad 
 N_i=e^{\rho} \check{N}_i\,, \quad
 \check{N}^i := \check{h}^{ij} \check{N}_i = e^\rho N^i\,,
\end{eqnarray}
we factorize the scale factor from the metric as
\begin{align}
 \dd s^2&=e^{2\rho}  [ - (N^2 - \check{N}_i \check{N}^i) \dd \eta^2
 \cr & \qquad \qquad \qquad
 + 2 \check{N}_i \dd \eta \dd x^i + \check{h}_{ij} \dd x^i \dd x^j
		     ]\,. \label{Exp:ADMmetric}
\end{align}
Expanding the perturbations, 
${\cal Q}=\delta N(:=N-1), \check{N}_i, \zeta$, and 
$\delta \gamma_{ij}$ as
${\cal Q} = {\cal Q}_1 + {\cal Q}_2 + \cdots$,
the zeroth-order Hamiltonian constraint equation yields the
background Friedmann equation:
\begin{eqnarray}
 6 \rho'\,^2 = \phi'\,^2 + 2 e^{2\rho} V(\phi)~, \label{Eq:Friedmann}
\end{eqnarray}
where a prime ``$~{}'~$'' denotes the differentiation 
with respect to the conformal time $\eta$.
The constraint equations at the linear order are obtained as
\begin{eqnarray}
 && e^{2\rho} V \delta N_1 - 3 \rho' \zeta'_1 
   + \partial^2  \zeta_1 + \rho' \partial^i \check{N}_{i,1} =0~,
 \label{Eq:Hconst1}\\
 && 4 \partial_i \left( \rho' \delta N_1 - \zeta'_1 \right)
    - \partial^2 \check{N}_{i,1} 
 + \partial_i \partial^j \check{N}_{j,1}=0~, \label{Eq:Mconst1}
\end{eqnarray}
where $\partial^2:=\partial^i \partial_i$. 
The higher-order constraints
can be obtained similarly.

\subsection{Residual gauge degrees of freedom}  
\label{SSec:originIR}
The constraint equations (\ref{Eq:Hconst}) and (\ref{Eq:Mconst}) allow us to
describe the non-dynamical variables $N$ and $N_i$ in terms of
$\zeta$. Here we stress that the constraints (\ref{Eq:Hconst1}) and
the divergence of (\ref{Eq:Mconst1}) are elliptic-type equations, 
which require boundary conditions to solve. Even though we impose 
the gauge conditions (\ref{comoving}) and (\ref{Cond:comoving}) 
at each space-time point, $N$ and $N_i$ are not uniquely determined 
because of the presence of such non-local gauge degrees of freedom. 
At the first order of perturbation, these degrees of freedom
are studied in Ref.~\cite{IRgauge_L}, where general solutions of 
$\delta N_1$ and $N_{i,1}$ are given in the form:
\begin{align}
 & \delta N_1 = \frac{1}{\rho'} \left( \zeta_1' - \frac{1}{4}
  \partial^i G_{i}  \right)~,  \label{Exp:N1} \\
 & \check{N}_{i,1}= \partial_i 
 \left( \frac{\phi'\,^2}{2\rho'\,^2} \partial^{-2} \zeta_1'
 - \frac{1}{\rho'}  \zeta_1 \right) \cr
& \qquad \qquad
 - \frac{1}{4} \left( 1 + \frac{\phi'\,^2}{2 \rho'\,^2} \right) \partial_i
 \partial^{-2} \partial^j G_{j}
 +G_i~. \label{Exp:cNi1}
\end{align}
Here, an arbitrary vector function $G_{i}(x)$ that satisfies the Laplace
equation $ \partial^2 G_{i}(x) =0$ was introduced to make 
explicit the presence of degrees of freedom corresponding to the 
boundary conditions. Substituting
Eqs.~(\ref{Exp:N1}) and (\ref{Exp:cNi1}) into the equations of motion
for $\zeta_1$ and $\delta \gamma_{ij,1}$, we find that 
the introduction of the gauge function $G_i(x)$ modifies 
their evolution equations as well~\cite{IRgauge_L}. 

The ambiguity originating from the choice of the vector $G_i(x)$
is a sign of the presence of residual gauge degrees of freedom.
Here, we explicitly show that $G_i(x)$ 
represents the residual gauge degrees of freedom that 
remain undetermined even after specifying the gauge by 
the conditions~(\ref{comoving}) and 
(\ref{Cond:comoving}).
Since the gauge condition $\delta \phi=0$ completely fixes the
temporal gauge, the residual gauge can reside only in 
changing the spatial coordinates:
$x^i \to \tilde{x}^i=x^i+\delta x^i$. 
The metric perturbations then transform as 
\begin{align}
 & \tilde{\check{N}}_{i,1}(x) = \check{N}_{i,1}(x) - \delta x_i'~, \label{Trans:cNi1} \\
 & \tilde{\zeta}_1(x) = \zeta_1(x) - \tfrac{1}{3} \partial^i \delta
 x_i~, \label{Trans:zeta1} \\
 & \delta \tilde{\gamma}_{ij,1}(x) = \delta \gamma_{ij,1}(x) -2 \left( \partial_{(i} \delta x_{j)}  
 - \tfrac{1}{3} \partial^k \delta x_k \delta_{ij} \right)~. \label{Trans:gamma1}
\end{align}
In this section 
we associate a tilde ``$~\tilde{}~$'' with the 
perturbed variables in the gauge 
with $G_{i}\ne 0$ to discriminate them 
from the perturbed variables in the gauge with $G_{i} = 0$.

Since we have not changed the temporal coordinate, the lapse function
remains unchanged at the linear order. 
Equating $\delta N_1$ with $\delta \tilde{N}_1$,
given by Eq.~(\ref{Exp:N1}), we find that $\zeta_1$ is
related to $\tilde{\zeta}_1$ as
\begin{eqnarray}
 \tilde{\zeta}'_1 = \zeta'_1 + \frac{1}{4} \partial^i G_i ~.
 \label{Trans:zeta1/2}
\end{eqnarray}
Comparing Eq.~(\ref{Trans:zeta1}) with (\ref{Trans:zeta1/2}), we obtain
\begin{eqnarray}
 \partial^i \delta x_i' = - \tfrac{3}{4} \partial^i  G_i~. 
 \label{Exp:deltaxtrace}
\end{eqnarray}
Imposing the transverse condition on $\delta \gamma_{ij}$ in
Eq.~(\ref{Trans:gamma1}), we obtain another condition for $\delta x^i$ as
\begin{eqnarray}
 \partial^2 \delta x_i = - \tfrac{1}{3} \partial_i \partial^j \delta
  x_j~. \label{Cond:deltaxi}
\end{eqnarray}
Then, Eqs.~(\ref{Exp:deltaxtrace}) and (\ref{Cond:deltaxi}) are
integrated to give
\begin{align}
 \delta x_i& = - \int \dd \eta \, G_i(x)+\frac{1}{4} \int \dd \eta\, \partial_i \partial^{-2}
  \partial^j G_j(x) \cr &\qquad \quad  
 +  \!\int\! \dd \eta \, h_i(x) + H_i(\bm{x}), 
\end{align}
where we introduced vector functions $h_i(x)$ and $H_i(\bm{x})$ 
that satisfy
\begin{align}
 &\partial^i h_i(x)= \partial^2 h_i(x)=0~, \\
 & 3 \partial^2 H_i(\bm{x}) + \partial_i \partial^j H_j(\bm{x})=0~.
\end{align}
Using Eqs.~(\ref{Trans:zeta1}) and (\ref{Exp:deltaxtrace}),
Eq.~(\ref{Exp:cNi1}) is recast into 
\begin{align}
  \tilde{\check{N}}_{i,1}(x)& = \check{N}_{i,1}(x) 
 + \frac{1}{3 \rho'} \partial_i \partial^j \delta x_j \cr & \qquad \quad
 - \frac{1}{4} \partial_i \partial^{-2} \partial^j G_{j,1}(x) +
 G_{i,1}(x)~.  \label{Trans:TNi1/ap2}
\end{align}
Comparing Eq.~(\ref{Trans:TNi1/ap2}) with Eq.~(\ref{Trans:cNi1}), the
vector function $h_i(x)$ is determined by
\begin{eqnarray}
 h_i(x) = \frac{1}{4\rho'} \int \dd \eta \, \partial_i \partial^j
  G_j(x) + \frac{1}{\rho'} \partial^2 H_i(\bm{x})~.
\end{eqnarray}
The degrees of freedom in boundary conditions are then found to
represent the change of the spatial coordinates:
\begin{align}
 \delta x_i(x) &=   -   \int \hspace{-0.1cm} \dd \eta\, G_{i}(x) +
 \frac{1}{4} \int \hspace{-0.1cm} \dd \eta \partial_i \partial^{-2}
  \partial^j G_{j}(x) \cr &\qquad 
 +  \frac{1}{4} \int\hspace{-0.1cm} \frac{\dd \eta}{\rho'} \int \hspace{-0.1cm} \dd \eta \, \partial_i
 \partial^j G_{j}(x)  + H_{i}(\bm{x}) 
 \cr &\qquad  + 
 \int \frac{\dd \eta}{\rho'} \, \partial^2 H_{i}(\bm{x})~, \label{Exp:dxi}
\end{align}
which is basically expressed in terms of $G_i(x)$. 
We note that the time-independent vector $H_i(\bm{x})$ 
can be absorbed into the integration constant 
of the temporal integral, 
$\int \dd\eta\, \{-G_i(x)+(1/4)\partial_i\partial^{-2}\partial^j G_j(x)\}
$. 
Substituting Eq.~(\ref{Exp:dxi}) into Eqs.~(\ref{Trans:zeta1}) and 
(\ref{Trans:gamma1}), we find that 
spatial components of metric perturbation
transform as
\begin{eqnarray}
 &&\!\!\!\!\!\!
 \tilde{\zeta}_1=\zeta_1 + \frac{1}{4}\hspace{-0.1cm}\int \hspace{-0.1cm} \dd \eta\,
 \partial^i G_{i} - \frac{1}{3} \partial^i H_{i}~. \\
 &&\!\!\!\!\!\!
 \delta \tilde{\gamma}_{ij,1} = \delta \gamma_{ij,1} + 
 \hspace{-0.1cm}\int \hspace{-0.1cm} \dd \eta \left\{ 2\partial_{(i} G_{j)}
  - \tfrac{1}{2} (\partial_i \partial_j \partial^{-2}+\delta_{ij})
 \partial^k G_{k}   \right\} \cr
 && \qquad  -\! \frac{1}{2}\hspace{-0.1cm} \int\hspace{-0.1cm}
 \frac{\dd \eta}{\rho'}\hspace{-0.1cm} \int\hspace{-0.1cm} \dd \eta \partial_i
 \partial_j \partial^k G_{k}\! - 2 \left\{ \partial_{(i} H_{j)}
 - \tfrac{1}{3} \partial^k H_{k} \delta_{ij} \right\} \cr
&&  \qquad  - 2 \int \frac{\dd \eta}{\rho'} \partial^2 \partial_{(i}
 H_{j)}.
\end{eqnarray}

When we consider the universe with infinite volume and require that all
quantities are regular at the spatial infinity, the solution of 
$\delta N_1$ and $\check N_{i,1}$ would be specified uniquely. However, in this
case, we observe the singular behaviour in the loop corrections of the
curvature perturbation $\zeta$~\cite{Sloth:2006az, Sloth:2006nu, Seery:2007we, Seery:2007wf, Urakawa:2008rb, Cogollo:2008bi, Rodriguez:2008hy, Seery:2009hs, Gao:2009fx, Giddings:2010nc, Bartolo:2010bu, Seery:2010kh, Kahya:2010xh}. 
This is because the IR fluctuation acausally propagates 
through the non-physical gauge modes and comes into play. 
In contrast, if we do not care about any
singular behaviors at infinity, that would never be observed by us, 
a variety of homogeneous solutions $G_i$ can be added to the solution of
$\delta N_1$ and $\check N_{i,1}$. In our
previous work~\cite{IRsingle}, we have 
shown that, by choosing the function $G_i$ appropriately, we can
guarantee the regularity of fluctuations of $\zeta$ in the flat gauge 
as long as a finite spatial region of our universe is concerned. 
We think that this is a remarkable progress, but 
the prescription given in Ref.~\cite{IRsingle} is 
not completely satisfactory in that 
the loop corrections for $\zeta$ depend on the choice of 
the boundary conditions for the lapse function and the shift vector. 
This fact signifies that the $n$-point functions for the
``so-called'' curvature perturbation $\zeta$ is not a 
genuine gauge-invariant quantity,
when we take into account 
the gauge degrees of freedom associated with the 
choice of boundary conditions. 
Our discussion here can be extended straightforwardly 
to the higher-order in perturbations.

\section{Gauge-invariant quantities} \label{Sec:GIQ}
In this section, we provide one simple example of genuine
gauge-invariant quantities. If we compute genuine 
gauge-invariant quantities, the results by definition should be
unaffected by the choice of the gauge. Hence, they should be IR regular 
even if we calculate them based 
on the standard perturbation theory. 
We demonstrate this in the following two sections.

\subsection{Definitions of scalar curvatures} \label{SSec:Def}
One simple way to realize the gauge invariance is to use variables
defined in a completely fixed slicing and threading. 
What is revealed in the previous section is the fact that
the genuine gauge-invariant variables cannot be constructed by 
simply adapting gauge conditions to metric components 
at each space-time point. In order to fix the gauge completely, 
we also need to remove
the unphysical degrees of freedom associated with the choice of boundary
conditions. This cannot be achieved easily due to the difficulties in
removing all arbitrariness regarding the choice of space-time 
coordinates. It is, however, possible to calculate 
genuine gauge-invariant quantities
even if we do not accomplish the complete gauge fixing.   

Since the time slicing is uniquely fixed by the gauge
condition $\delta \phi=0$, it is enough if we can arrange
quantities so as to be invariant under the transformation 
of spatial coordinates. 
In our previous work~\cite{IRgauge_L}, we proposed to calculate
$n$-point functions for the scalar curvature of the induced metric on 
a $\phi\!=$constant surface, $^s\!R$. 
Although $\sR$ itself does not remain invariant but 
transforms as a scalar quantity under the change of spatial
coordinates, the gauge invariance of the $n$-point functions 
of $\sR$ would be ensured, if we could specify its $n$ arguments in a
coordinate-independent manner. The distances of spatial geodesics 
that connect pairs of $n$ points characterize the configuration 
in a coordinate independent manner. 
Based on this idea, we specify the $n$ spatial points in terms of the 
geodesic distances and the directional cosines, measured from 
a reference point. 
Although we cannot specify the reference 
point and frame in a coordinate independent manner, 
this gauge dependence would not matter as long as we are interested in 
the correlation functions in a quantum state that respects the spatial 
homogeneity and isotropy of the universe. 

We consider the three-dimensional geodesics whose affine parameter ranges
from $\lambda=0$ to $1$ with the initial ``velocity'' given by
$$
\left.{\dd x^i(\bm{X},\lambda)
\over \dd \lambda}\right\vert_{\lambda=0}= X^i~.
$$   
We identify a point in the geodesic normal coordinates $X^i$ 
with the end point of the geodesic 
$x^i(\bm{X},\lambda=1)$. 
Noticing that in the absence of the fluctuations $X^i$
coincides with $x^i$,  
we expand $x^i(\bm{X})$ as 
$$x^i(\bm{X}):=X^i+\delta x^i(\bm{X})~.$$ 
We denote the spatial curvature whose argument is specified by 
the geodesic normal coordinates $X^i$ as 
\begin{align}
 \!\! ^g\! R(\eta,\,\bm{X}):= {^s\! R} (\eta,\,x^i(\bm{X}))~. 
\end{align}
Then, ${\gR}$ can be expanded as
\begin{align}
&\!\! ^g\! R(\eta,\,\bm{X})  \cr
&\, = \sum_{n=0}^\infty \frac{\delta x^{i_1} \cdots \delta x^{i_n}}{n!}
 \partial_{i_1} \cdots \partial_{i_n}\!{^s\!R}(\eta,\, x^i)\vert_{x^i=X^{(i)}}\,. 
\label{Def:Rg}
\end{align}
The $n$-point functions of ${\gR}$ would be surely gauge-invariant, 
unless
the initial quantum state breaks the gauge-invariance.

Our main purpose of this paper is to demonstrate the
absence of IR divergence in the genuine gauge-invariant quantities at
one-loop order. 
At this order, the following three terms contribute to
the two-point function: 
\begin{eqnarray}
 \langle {^g\!R} {^g\!R} \rangle_{4} :=
  \langle {^g\!R}_1 {^g\!R}_3 \rangle +  \langle {^g\!R}_2 {^g\!R}_2 \rangle
  + \langle {^g\!R}_3 {^g\!R}_1 \rangle~, \label{Exp:1loop}
\end{eqnarray}
where the subscripts $1,2,3,4$ mean the numbers of the contained 
creation and annihilation 
operators or equivalently the number of the contained 
interaction picture field operators. 
For simplicity, we neglect the terms that do not
yield the IR divergence. In the above expression, each term 
contains two pairs of contraction between creation and annihilation 
operators. Only when one of these pairs does not contain any
differentiation, the term potentially contributes to IR divergence. 
Since the loop integrals diverge at most 
logarithmically, one spatial or temporal derivative is
sufficient to remedy their divergent behaviors. 
Noticing that the curvature
perturbation in the first-order scalar curvature is multiplied by the
spatial derivatives as
\begin{eqnarray}
 \gR_1 = \sR_1 \propto \partial^2 \zeta_1~,
\end{eqnarray}
the terms in $\gR_3$ that include more
than one interaction picture field operators 
with spatial or temporal derivatives do not yield
divergences. 
This statement also applies to the terms in
$\gR_2$. 
Since $\gR_2$ contains at least 
one interaction picture field operator that is differentiated, 
the terms in $\gR_2$ that include more than one differentiated 
interaction picture field operators do not yield IR divergences.

The loops of gravitational wave perturbation 
without derivatives yield the logarithmic 
divergence, too. However, the gravitational wave perturbation $\delta\gamma_{ij}$
with derivatives no longer contributes to such 
divergent loop corrections.

Hereafter, we denote an equality 
which is valid only when we neglect 
the terms irrelevant to IR divergences by
``${\Approx}$''.  
Then, abbreviating 
the unimportant pre-factor, we simply denote the scalar
curvature $\sR$ as  
\begin{equation}
 \sR \Approx e^{-2\zeta} \left[ e^{-\delta \gamma} \right]^{ij}
 \partial_i \partial_j\zeta.  \label{Def:sR}
\end{equation}

\subsection{Gauge transformation} 
\label{SSec:tranformation}
To calculate the non-linear corrections under the slow-roll
approximation, it is convenient to temporally 
work in the flat gauge:
\begin{eqnarray}
 \tilde{h}_{ij}= e^{2\rho} \left[ e^{\delta \tilde{\gamma}} \right]_{ij}, \quad
 {\rm tr} [\delta \tilde{\gamma}]=0, \quad \partial^i \delta
 \tilde{\gamma}_{ij}=0, \label{Exp:metricflat}
\end{eqnarray}
because all the interaction vertexes are explicitly suppressed by the
slow-roll parameters in this gauge~\cite{Maldacena2002, IRsingle}. 
Here in this section we associate a tilde with the
metric perturbations in the flat gauge to discriminate those in the
comoving gauge.
The action in this gauge is given by
\begin{align}
 S&\Approx \frac{\Mp^2}{2} \!\int\! e^{2\rho} \Bigl[ \tilde{N}^{-1} 
  \left( \phi' + \varphi' - \tilde{\check{N}}^i \partial_i \varphi
 \right)^2 \cr & \qquad \qquad \quad 
 - 2 \tilde{N} e^{2\rho}\!\sum_{m=0} \frac{V^{(m)}}{m!}\varphi^m
 - \tilde{N} \tilde{\check{h}}^{ij} \partial_i\varphi
 \partial_j \varphi \cr & \qquad \qquad \quad 
 + \tilde{N}^{-1} \left( - {\rho'}^2 + 4 \rho' \partial_i \tilde{\check{N}}^i \right)  \Bigr]\! \dd \eta \dd^3 \bm{x}.
 \label{Exp:S/flat}
\end{align}

The transformation formulae between the comoving gauge and the flat gauge
are studied in Ref.~\cite{Maldacena2002}, and we briefly summarize them
in Appendix~\ref{Sec:transformation}. The curvature perturbation in 
the comoving gauge $\zeta$
is related to the fluctuation of the dimensionless scalar field (divided
by $\Mp$) in the flat gauge $\varphi$ as
\begin{align}
 \zeta& \Approx \zeta_n
   +\zeta_n\partial_\rho\zeta_n 
   +\frac{\varepsilon_2}{4} \zeta_n^2
  +{\zeta_n^2 \partial_\rho^2\zeta_n\over 2} \cr & \qquad \qquad 
  +{3 \varepsilon_2\zeta_n^2 \partial_\rho \zeta_n\over 4} 
  + \frac{1}{12} \varepsilon_2 (\varepsilon_2 +
 2\varepsilon_3)\zeta_n^3, \label{Exp:zeta0}
\end{align}   
where we have introduced $\zeta_n:=- (\rho'/ \phi') \varphi$, 
following Ref.~\cite{Maldacena2002}. We use the horizon flow
function: 
\begin{eqnarray}
 \varepsilon_0 := \frac{H_i}{H}~, \quad
 \varepsilon_{m+1} := \frac{1}{\varepsilon_m} \frac{\dd
 \varepsilon_m}{\dd \rho}~\qquad \quad {\rm for}~m \geq 0,  \label{Def:HFF}
\end{eqnarray}
where $H$ is the Hubble parameter and $H_i$ is the one
at the initial time.
The horizon flow function is related to the conventional slow-roll
parameters as shown in Ref.~\cite{Schwarz:2001vv}. Hereafter, assuming that the
horizon flow functions $\varepsilon_m$ with $m \geq 1$ are
all small of ${\cal O}(\varepsilon)$, we neglect 
the terms of ${\cal O}(\varepsilon^3)$.
In Eq.~(\ref{Exp:zeta0}), we neglected the cubic terms that include
only one graviton field $\delta \tilde{\gamma}_{ij}$, for the following reason. 
Since $\gR_1$ includes only $\zeta_1$, the terms in $\gR_3$ that
include only one graviton field $\delta \tilde{\gamma}_{ij}$, 
does not contribute to 
$\langle \gR \gR \rangle_4$ after taking the contraction. 

In line with the preceding papers~\cite{IRsingle, IRmulti}, 
in order to calculate the $n$-point functions, we solve the evolution
equation (Heisenberg equation) for the operator $\varphi$, 
and we express $\varphi$ in terms of the interaction picture field 
$\varphi_I$.  
Variation of the total action with respect to $\varphi$
yields 
\begin{align}
 & e^{-2\rho}  \partial_\eta
 \left[ \frac{e^{2\rho}}{\tilde{N}} \left(\phi' + \varphi' \right)
 \right] + \tilde{N} e^{2\rho} \sum_{m=0} \frac{V^{(m+1)}}{m!} \varphi^{m} \cr
 & \quad - \left( \phi' + \varphi' \right) \frac{1}{\tilde{N}}
 \partial_i \tilde{\check{N}}^i - \tilde{N}  \left[ e^{-\delta \tilde{\gamma}}
			       \right]^{ij} 
 \partial_i \partial_j \varphi  \Approx 0~,  \label{Eq:flat}
\end{align}
where $V^{(m)}:= \dd^m V/ \dd \phi^m$. 
To address the regularity of the graviton loops, we also include the
contributions from the gravitational wave perturbation. 
Variations with
respect to the lapse function and the shift vector, respectively, yield the
Hamiltonian constraint:
\begin{align}
 &(\tilde{N}^2-1) e^{2\rho} V  +  \tilde{N}^2 e^{2\rho} \sum_{m=1}^\infty
 \frac{V^{(m)}}{m!} \varphi^m 
\cr & \qquad \qquad  \qquad 
  + 2 \rho' \partial_i \tilde{\check{N}}^i +  \phi' \varphi'
  + \frac{1}{2} {\varphi'}^2 \Approx 0\,,
\end{align}
and the momentum constraints:
\begin{eqnarray}
 2\rho' \partial_i \tilde{N} - \tilde{N}(\phi' \partial_i \varphi + \partial_i \varphi
  \varphi') \Approx 0\,.
\end{eqnarray}
For the calculation of one loop corrections, it is enough to solve the
constraint equations up to the quadratic order. These constraint equations are solved to give
\begin{eqnarray}
  \delta \tilde{N}& \Approx&  -{{\phi'}^2\over
   2{\rho'}^2}\zeta_n+{1\over 4 \rho'}\varphi \left(
   \phi' \delta \tilde{N}_1 + \varphi' \right)\cr
  & \Approx&  -\varepsilon_1 \zeta_n+ \frac{\varepsilon_1}{2} \left(
							 \varepsilon_1 +
							\frac{\varepsilon_2}{2}\right)
  \zeta_{n}^2, \label{Exp:N/flat} \\ 
   \partial_i \tilde{\check{N}}^i& \Approx& \varepsilon_1 \zeta'_n
  -\frac{1}{2} \varepsilon_1 \varepsilon_2 \zeta_n \zeta'_n. \label{Exp:Ni/flat}
\end{eqnarray}
Substituting Eqs.~(\ref{Exp:N/flat}) and (\ref{Exp:Ni/flat}) into
Eq.~(\ref{Eq:flat}), the evolution equation of $\zeta_n$ is recast into 
a rather compact expression, 
\begin{eqnarray}
 {\cal L} \zeta_n&\Approx&  
 \left[ -2\varepsilon_1 \zeta_n + \frac{1}{2}
			      \varepsilon_1 (4\varepsilon_1 +
			      \varepsilon_2) \zeta_n^2 \right] \frac{1}{{\rho'}^2}
   \partial^2 \zeta_n 
\cr && \quad 
-  \varepsilon_1 \varepsilon_2  \zeta_n
  \partial_\rho \zeta_n - \frac{3}{4}  \varepsilon_2
  \varepsilon_3  \zeta_n^2 
\cr &&\quad
  + \left( \left[ e^{-\delta
					  \tilde{\gamma}}\right]^{ij} -
  \delta^{ij} \right)  \frac{1}{{\rho'}^2} \partial_i \partial_j \zeta_n
  \label{Eq:flat2}
\end{eqnarray} 
where the differential operator ${\cal L}$ is defined by 
\begin{equation}
{\cal L} :=  \partial_\rho^2 +  (3 - \varepsilon_1+ \varepsilon_2) \partial_\rho
 - \frac{1}{{\rho'}^2} \partial^2.
\end{equation}

We expand $\zeta_n$ as 
$\zeta_n= \zeta_{n,1} + \zeta_{n,2} + \zeta_{n,3} + \cdots$ and denote
$\zeta_{n,1}$ simply as $\psi:=\zeta_{n,1}$. The equation of motion
(\ref{Eq:flat2}) is expanded as
\begin{align}
 & \quad {\cal L} \psi = 0~, \\
 & {\cal L} \zeta_{n,2} \Approx - \varepsilon_1 \varepsilon_2 \psi
 \partial_\rho \psi - \frac{3}{4} \varepsilon_2 \varepsilon_3 \psi^2 
  \cr & \qquad \qquad 
 - 2 \varepsilon_1 \psi \frac{1}{{\rho'}^2} \partial^2 \psi
 -  \frac{1}{{\rho'}^2} \delta \tilde{\gamma}^{ij}_1 \partial_i
 \partial_j \psi~, \label{Eq:zetan2} \\
 & {\cal L} \zeta_{n,3} \Approx -  \frac{2}{{\rho'}^2} \varepsilon_1
  (\psi \partial^2 \zeta_{n,2} + \zeta_{n,2} \partial^2 \psi)  \cr
 & \qquad \qquad   +
 \frac{1}{2 {\rho'}^2} \varepsilon_1 (4 \varepsilon_1 + \varepsilon_2) \psi^2
 \partial^2 \psi -  \frac{1}{{\rho'}^2} \delta \tilde{\gamma}^{ij}_1
 \partial_i \partial_j \zeta_{n,2} \cr 
  & \qquad \qquad   +  \frac{1}{2 {\rho'}^2} (\delta \tilde{\gamma}_1^2)^{ij} 
     \partial_i \partial_j \psi~. \label{Eq:zetan3}
\end{align}
Here, we neglected $\delta \tilde\gamma_{ij,2}$ on the
right-hand side of Eq.~(\ref{Eq:zetan3}), 
because a particular solution of
$\delta \tilde\gamma_{ij,2}$ is associated with derivatives, 
and we set its
homogeneous solution to zero. 
At the second order, Eq.~(\ref{Eq:zetan2}) is
integrated to give
\begin{align}
 \zeta_{n,2}
   \Approx \breve{\zeta}_{n,2} + \frac{1}{2} \delta \tilde\gamma^{ij}_1
  x_i \partial_j \psi, \label{Exp:zetan2}
\end{align}
where, for a later use, we have distinguished the part containing 
the contributions due to gravitational waves 
from the pure scalar part $\breve{\zeta}_{n,2}$ given by
\begin{align}
 \breve{\zeta}_{n,2}& \Approx \left( \frac{\varepsilon_1}{2} + \xi_2 \right) \psi^2
   + \varepsilon_1  \psi \partial_\rho\psi 
   + \varepsilon_1 (\varepsilon_1 + \varepsilon_2) \psi \partial_\rho
 \psi \cr &\qquad \qquad +\delta\zeta_{n,2} + \lambda_2
 \psi(\partial_\rho -x^i \partial_i) \psi~. \label{Sol:zetan2}
\end{align}
Here, $\breve{\zeta}_{n,2}$ includes the non-local term:
\begin{equation}
  \delta\zeta_{n,2}:=-{\cal L}^{-1}
  \left[ \frac{3}{4}  \varepsilon_2 (2 \varepsilon_1 +
   \varepsilon_3) \psi^2 \right]~. 
\end{equation}
It should be emphasized that the homogeneous solutions $\xi_2 \psi^2$
and $\lambda_2 \psi(\partial_\rho -x^i \partial_i) \psi$ can be added to
$\breve{\zeta}_{n,2}$, where the time dependent functions $\xi_2$ and $\lambda_2$
should be of ${\cal O}(\varepsilon^2)$ and their derivatives should be of
${\cal O}(\varepsilon^3)$. One can easily check that 
the above solution satisfies Eq.~(\ref{Eq:zetan2}) 
to the present order of approximation, 
using the commutation relations
\begin{eqnarray}
 && \left[ {\cal L},\, \partial_\rho \right]= -2(1-\varepsilon_1)
  \frac{1}{{\rho'}^2}  \partial^2 + {\cal O} (\varepsilon^2)~, \cr
 && \left[ {\cal L},\, x^i \partial_i \right] = - \frac{2}{{\rho'}^2} 
 \partial^2~, \\
 && \left[ {\cal L},\, 1/{\rho'}^2 \right] = - \frac{2}{{\rho'}^2}
  (2 \partial_\rho + 1)  + {\cal O} (\varepsilon)~.   \nonumber
\end{eqnarray}
We are also allowed to change the solution of $\breve{\zeta}_{n,2}$ at
${\cal O}(\varepsilon)$ by adjusting its solution at 
${\cal O}(\varepsilon^2)$\footnote{Actually, we could add a term of ${\cal O}(\varepsilon)$ proportional to 
$\psi (\partial_\rho - x^i \partial_i) \psi$ to $\breve\zeta_{n,2}$.}. In the succeeding section, we will explain
that the solution of $\breve{\zeta}_{n,2}$ is restricted by the
requirement that the canonical commutation relation should be
consistently satisfied. This requirement is, however, not enough to determine
$\check{\zeta}_{n,2}$ uniquely. Therefore, in Eq.~(\ref{Sol:zetan2}), we
fixed the terms of ${\cal O}(\varepsilon)$, requesting that, in addition to the
consistency of the commutation relation, $\check{\zeta}_{n,2}$ should be
kept in the simplest form.

At the third-order of perturbation, Eq.~(\ref{Eq:zetan3}) is integrated to give
\begin{align}
  \zeta_{n,3} \Approx \breve{\zeta}_{n,3} +\frac{1}{8}\left( \delta \tilde{\gamma}_1 \delta
			   \tilde{\gamma}_1 \right)^{ij} x_i \partial_j
 \psi  + \frac{1}{8} \delta \tilde\gamma^{ij}_1 \delta \tilde\gamma^{kl}_1 x_j
 x_l \partial_i \partial_k \psi,  \label{Exp:zetan3}
\end{align}
where $\breve{\zeta}_{n,3}$ is the part purely composed of
the scalar perturbation as
\begin{align}
 \breve\zeta_{n,3}& \Approx \xi_3 \psi^3 +\lambda_3 
  \psi^2 (\partial_\rho - x^i \partial_i) \psi 
 \cr & \qquad  + \frac{1}{2} \varepsilon_1^2 \psi^2
  \partial_\rho^2 \psi + \frac{1}{4} \varepsilon_1 (6 \varepsilon_1 -
  \varepsilon_2) \psi^2 x^i \partial_i \psi .
\end{align}
It is again allowed to add homogeneous solutions whose coefficients 
$\xi_3$ and $\lambda_3$ are of
${\cal O}(\varepsilon^2)$ and their derivatives by $\rho$ is of 
${\cal O}(\varepsilon^3)$. Here we
note that $[{\cal L},\, \partial_\rho - x^i \partial_i]={\cal
O}(\varepsilon)$.

\subsection{Consistency of commutation relations}

We have to take into account the following additional 
conditions that determine the choice of the homogeneous
solution in $\zeta_n$. 
Until Eq.~(\ref{Def:xi3}), we will leave $\xi_3$ unspecified, 
but the other time dependent functions $\xi_2$, $\lambda_2$ and $\lambda_3$ 
are constrained in principle so as to guarantee the normal commutation
relation for $\psi$, as we will explain soon below. 

The evolution of the Heisenberg field $\zeta_n$ is usually
supposed to be solved with the initial conditions 
that the Heisenberg field $\zeta_n$ is 
identified with the interaction picture field $\psi$ at the initial time. 
This procedure guarantees that the operator $U$ that relates $\psi$
to $\zeta_n$ by $\psi=U\zeta_n U^\dag$ is unitary. 
In this case, the canonical commutation relation for $\zeta_n$ 
\begin{eqnarray}
 [\zeta_n(\eta, \bm{x}),\,\pi_n(\eta,
\bm{y})]=i\delta^{(3)}(\bm{x}-\bm{y}),  \label{Rel:Com}
\end{eqnarray}
is equivalent to the commutation relation for 
the interaction picture fields 
\begin{eqnarray}
 [\psi(\eta, \bm{x}),\,\pi_\psi(\eta,
\bm{y})]=i\delta^{(3)}(\bm{x}-\bm{y}), \label{Rel:Com_psi}
\end{eqnarray}
where $\pi_n$ is the conjugate momentum of $\zeta_n$ and 
$\pi_\psi$ is its linear truncation. 
We give a more explicit expression only 
up to ${\cal O}(\varepsilon)$ relative to the leading term here. 
In this approximation, using Eq.~(\ref{Exp:S/flat}), we obtain the
kinetic term in the action as
\begin{eqnarray*}
 S_{\rm kin} &\!\! 
    = \!\! & \int \dd\eta\int \dd^3x
    {\Mp^2 e^{2\rho}\over 2\tilde{N}}(\phi'+\varphi')^2+\cdots\cr
 &\!\!  \supset \!\!  &
   \int \dd\eta\int \dd^3x
   {\Mp^2 e^{2\rho}}\varepsilon_1\zeta'_n\cr
&& \quad\times
      \left[(1+\varepsilon_1 \zeta_n)\zeta'_n+
      \rho'(\varepsilon_2-2\varepsilon_1)\zeta_n+{\cal O}(\varepsilon^2)\right]. 
\end{eqnarray*}
From this expression, we can define the conjugate momentum 
\begin{eqnarray*}
\pi_n &\!\! :=\!\!  &{\delta S_{\rm kin}\over\delta\zeta'_n}\nonumber\\
  &\!\! =\!\! & \Mp^2 \varepsilon_1 e^{2\rho}
      \left[2(1+\varepsilon_1 \zeta_n)\zeta'_n+
      \rho'(\varepsilon_2-2\varepsilon_1)\zeta_n+{\cal O}(\varepsilon^2)\right]. 
\end{eqnarray*}

In the preceding subsection, we gave the non-linear solution 
by integrating the equation of motion without care about its
initial conditions. Therefore unitary relation between 
$\psi$ and $\zeta_n$ is not guaranteed. 
Once we obtain the definite expansions of $\zeta_n$ and
$\pi_n$ in terms of $\psi$ and $\pi_\psi$, it would be possible to check
whether the commutation relation of $\psi$ and $\pi_\psi$ is guaranteed from that of
$\zeta_n$ and $\pi_n$ or vise versa. 
We can here check this consistency of these commutation relations to the 
only limited extent because we have neglected the terms containing more than 
two interaction picture field operators with space-time
differentiation. Under this limitation, 
we can evaluate the commutator, assuming 
$[\psi(\eta, \bm{x}),\,\pi_\psi(\eta,
\bm{y})]=i\delta^{(3)}(\bm{x}-\bm{y})$, as 
\begin{align}
 &[\zeta_n(\eta, \bm{x}),\,\pi_n(\eta,
\bm{y})] \cr
 &= i\delta^{(3)}(\bm{x}-\bm{y}) \Bigl[1+\left\{4\xi_2-3\lambda_2+{\cal
 O}(\varepsilon^2)\right\}\zeta_n  \cr 
 & \qquad \qquad \qquad \quad +
    \left\{6\xi_3 +3 \lambda_3 +{\cal O}(\varepsilon^2)\right\}\!\zeta_n^2+\cdots
    \Bigr], 
\label{commutator}
\end{align}
where the ellipsis represents the 
terms containing $\pi_\psi$ and spatial derivatives, 
which are the beyond the scope of the present paper.   
In this way we can verify that the solution we gave 
is consistent with the expected commutation relation 
at ${\cal O}(\varepsilon)$. If we have chosen 
an inappropriate solution for $\check{\zeta}_{n,2}$ at ${\cal
O}(\varepsilon)$, the commutation relation would not be satisfied. 
  
At ${\cal O}(\varepsilon^2)$
the terms 
that we could evaluate in (\ref{commutator}) 
include the unspecified functions $\xi_2$,
$\lambda_2$, and $\lambda_3$. From the requirement that the right-hand
side of Eq.~(\ref{commutator}) should be equated to Eq.~(\ref{Rel:Com}),
$\lambda_2$ and $\lambda_3$ are related to $\xi_2$ and $\xi_3$, respectively.

\subsection{Calculations of scalar curvature} 
In this subsection, we give the expansion of the scalar curvature 
of a $\phi=$constant hypersurface 
$\gR$ in terms of $\psi$, which is the interaction picture field of
$\zeta_n$. Using Eqs.~(\ref{Exp:zetan2}) and (\ref{Exp:zetan3}), 
we first perturb the scalar curvature $\sR$, given by
Eq.~(\ref{Def:sR}), to obtain
\begin{align}
 \label{Exp:sR1}
 \sR_1 & = \sbR_1=  \partial^2\psi, \\
  \label{Exp:sR2} 
 \sR_2 & \Approx  \sbR_2 + \frac{1}{2} \delta \gamma_1^{ij} x_i
 \partial_j\partial^2 \psi , \\
 \sR_3 & \Approx \sbR_3 + \frac{1}{8}\delta \gamma^{ij}_1 \delta \gamma^{kl}_1
x_j x_l \partial_i \partial_k \partial^2 \psi 
  + \frac{1}{8} (\delta \gamma_1^2)^{ij} x_i \partial_j \partial^2 \psi
 , \nonumber \\ \label{Exp:sR3} 
\end{align}
where $\sbR_2$ and $\sbR_3$ are defined by
\begin{align}
 \sbR_2 & \Approx \partial^2 \breve\zeta_{n,2} + \psi 
     \,\partial^2\!\left( \partial_\rho -2 + \varepsilon_2/2 
 \right) \psi, \label{Exp:sbR2} \\
 \sbR_3 & \Approx  \partial^2 \breve\zeta_{n,3} + \breve\zeta_{n,2} 
     \partial^2 \left( \partial_\rho -2 + \varepsilon_2/2 \right)
 \partial^2 \psi \cr
 & \qquad   + \psi \left( \partial_\rho -2 + \varepsilon_2/2
 \right) \partial^2 \breve\zeta_{n,2} + \partial^2 \psi \partial_\rho \breve\zeta_{n,2} \cr
&  \qquad + {\psi^2\over 2} \,\partial^2\! 
  \Bigl[  \partial^2_\rho -
  4(\partial_\rho - 1)  + \frac{3}{2} \varepsilon_2 \partial_\rho \cr & \qquad \qquad \qquad \qquad  - 3 \varepsilon_2 + \frac{1}{2}\varepsilon_2
 (\varepsilon_2 + \varepsilon_3) \Bigr] \psi.  \label{Exp:sbR3}
\end{align}
To derive Eqs.~(\ref{Exp:sR2}) and (\ref{Exp:sR3}), we have used 
the fact that,
as presented in Eq.~(\ref{Exp:Transgamma}), the gravitational wave perturbation in the comoving 
gauge $\delta \gamma_{ij}$ is identical to that in the flat gauge
$\delta \tilde\gamma_{ij}$, besides the terms irrelevant to IR divergences.  

Noticing that the spatial metric after removing the common scale factor is given by
\begin{eqnarray}
 \dd \lambda^2 = e^{2\zeta} \left[ e^{\delta \gamma} \right]_{ij} \dd
  x^i \dd x^j~,
\end{eqnarray}
the geodesic normal coordinates $X^i$ is given by 
\begin{equation}
x^i(\bm{X}) \Approx e^{-\zeta} {\bigl[ e^{- \delta \gamma/2} \bigr]}{}^{\!i}_j X^j, 
\end{equation}
where we again abbreviated the terms that include space-time derivatives. 
The difference between the global coordinates and the geodesic normal ones 
is given by 
\begin{eqnarray}
 \delta x^i&:=& x^i(\bm{X}) - X^i = \delta x^i_1+ \delta x^i_2 + \cdots
\end{eqnarray}
where 
\begin{align}
 \delta x^i_1 &\Approx - \psi X^i -\frac{1}{2}  \delta \gamma^{ij}_1 X_j~,
 \label{Exp:dx1} \\ 
 \delta x^i_2 &\Approx - \zeta_2 X^i  + \frac{1}{2} \psi^2 X^i + \frac{1}{8}
 (\delta \gamma_1^2)^{ij} X_j ~.  \label{Exp:dx2}
\end{align}

Now, we are ready to calculate the scalar curvature $\gR$.  
Substituting Eqs.~(\ref{Exp:sR2}) and (\ref{Exp:dx1}) into
Eq.~(\ref{Def:Rg}), we obtain
\begin{eqnarray}
 \gR_2 = \sR_2 + \delta x^i_1 \partial_i \sR_1 \Approx
 \sbR_2-\psi X^i\partial_i \sbR_1~, 
\label{Exp:gR2}
\end{eqnarray}
and substituting Eqs.~(\ref{Exp:sR3}) and (\ref{Exp:dx2}) into
Eq.~(\ref{Def:Rg}), we obtain
\begin{eqnarray}
 \gR_3 &=& \sR_3 + \delta x^i_1 \partial_i \sR_2 + \delta x^i_2
  \partial_i \sR_1 + \frac{1}{2} \delta x^i_1 \delta x^j_1
 \partial_i \partial_j \sR_1 \cr
  &\Approx& \sbR_3 - \psi X^i \partial_i \sbR_2 
 - \left( \breve{\zeta}_{n,2} + \frac{\varepsilon_2}{4} \psi^2 \right)X^i \partial_i
 \sbR_1 \cr && + \frac{1}{2} \psi^2 (X^i \partial_i)^2
 \sbR_1  \label{Exp:gR3}
\end{eqnarray}
It would be appropriate to emphasize that, in contrast to the
contributions from $\zeta$, the contribution from the gravitational 
wave perturbation $\delta \gamma_{ij}$ completely cancels with each
other in $\gR$. 
This clearly shows that the graviton loop does not lead to IR divergence at
the one-loop order. 
This is essentially because the effect on $\sR$ from IR modes of gravitational wave
perturbation is simply caused by the associated deformation of the spatial coordinates. 
Such gauge artifacts should completely disappear when we consider 
coordinate independent quantities like $\gR$.  


\section{IR regularity and gauge-invariant vacuum} 
\label{Sec:IRregularity}
This section is devoted to show how the possibly divergent terms are
cancelled in the $n$-point functions of $\gR$. As described in
Sec.~\ref{SSec:Def}, in the standard cosmological perturbation
the Hilbert space has not been reduced to the one 
that is composed only of the physical degrees of freedom. 
Namely, a part of gauge degrees of freedom are left unfixed. 
Hence, an arbitrary quantum state defined in this Hilbert space can 
be non-invariant along the gauge orbit of these residual gauge degrees 
of freedom. To ensure the
gauge-invariance of the $n$-point functions of $\gR$, it turns out 
to be crucial to set the initial quantum state to be gauge invariant as well. 
Otherwise, the $n$-point functions 
fail to be regular due to the gauge artifacts. In this section, we
reveal how the gauge-invariance condition(=regularity condition for the
$n$-point functions) restricts the initial vacuum, particularly
considering the one-loop correction to the two-point function.

\subsection{Proof of IR regularity}  
\label{SSec:Proof}
In the previous section, we expanded the scalar curvature $\gR$ in terms
of the interaction picture field $\psi$. 
We expand $\psi$ as
\begin{equation}
 \psi=\int{\dd^3 \bm{k} \over (2\pi)^{3/2}}\left(\psi_{\bk}a_{\bk}
   +\psi^*_{\bk}a^\dag_{\bk}\right), 
\end{equation}
where the creation and annihilation operators satisfy 
$ [a_{\bk},\, a^\dagger_{\bk'}]
=\delta^{(3)}(\bm{k}-\bm{k}')$. Focusing on the 
contribution from each Fourier mode $\psi_{\bk}=v_{\bk}e^{i\bk\bx}$, 
the derivative operator $x^i\partial_i$ is rewritten as
\begin{equation}
 x^i\partial_i\psi_{\bk}=v_{\bk}\partial_{\log k} e^{i\bk\bx}
 =\partial_{\log k} \psi - e^{i\bk\bx} \partial_{\log k} v_{\bk}.
\end{equation}

For illustrative purpose, we first consider the leading order in the
slow-roll approximation~\cite{IRgauge_L}. For the Bunch-Davies vacuum, the mode
function $v_k$ is given by
\begin{eqnarray}
 v_k(\eta) = - \frac{{\rho'}^2 e^{-\rho}}{\phi'} \frac{1}{k^{3/2}}
  \frac{i}{\sqrt{2}} e^{-ik\eta} (1+ ik \eta)~,  \label{Exp:vk_BD}
\end{eqnarray}
and is easily checked to satisfy
\begin{eqnarray}
 (\partial_\rho - x^i \partial_i) \psi_{\bk} = - D_k \psi_{\bk}~, \label{Exp:SI}
\end{eqnarray}
where the operator $D_k$ is defined by 
\begin{eqnarray}
  D_k:=\partial_{\log k}+ \frac{3}{2} ~. \label{Def:Dk}
\end{eqnarray}
Using Eq.~(\ref{Exp:SI}), $\gR_2$ and $\gR_3$  could be compactly written as
\begin{eqnarray}
 \gR_2 \Approx -\psi \partial^2 D_k\psi_{\bk}~, \qquad
 \gR_3 \Approx \frac{1}{2} \psi^2 \partial^2 D_k \psi_{\bk}~. 
\end{eqnarray}
Taking the contractions of $\gR$, we obtain
\begin{align*}
 & \langle {}\gR_3(X_1) \gR_1(X_2)\rangle \cr
 &  \Approx  \frac{1}{2} \langle \psi^2 \rangle \!
  \left[  
   \prod_{i=1,2} \int \frac{\dd^3 \bm{k}_i}{(2\pi)^{3/2}}
   \right]
    \left(D_{k_1}^2 k_1^2 \psi_{\bk_1}(X_1)\right)
 \cr & \qquad \qquad \qquad \qquad \qquad \times
 k_2^2 \psi^*_{\bk_2}(X_2) \delta^{(3)} \left( \bm{k}_1- \bm{k}_2 \right)   \cr
 & = \frac{1}{2} \langle \psi^2 \rangle \int \frac{\dd (\log\! k)}{2\pi^2}
 \partial_{\log k}^2 \left\{ k^{7/2} \psi_{\bk}(X_1) \right\}  k^{7/2} \psi^*_{\bk}(X_2),
\end{align*}
and
\begin{align*}
 & \langle {} \gR_2(X_1) \gR_2(X_2) \rangle \cr
 &  \Approx  \langle \psi^2 \rangle \!
  \left[  
   \prod_{i=1,2} \int \frac{\dd^3 \bm{k}_i}{(2\pi)^{3/2}}
   \right]
  \left( D_{k_1} k_1^2 \psi_{\bk_1}(X_1) \right)
  \cr & \qquad \qquad \qquad \qquad \qquad \times 
  \left(
   D_{k_2} k_2^2 \psi^*_{\bk_2}(X_2)\right)
    \delta^{(3)} \left( \bm{k}_1- \bm{k}_2 \right)   \cr
 & = \langle \psi^2 \rangle \!\!\int\! \frac{\dd (\log\! k)}{2\pi^2}
 \partial_{\log k}\! \left\{ k^{7/2} \psi_{\bk}(X_1) \right\}  \partial_{\log
 k}\! \left\{ k^{7/2} \psi_{\bk}(X_2) \right\}\!,
\end{align*}
where using the geodesic normal coordinate we defined
$X_m:=(\eta,\,\bm{X}_m)$ for $m=1,2$.
Gathering the three terms on the right-hand side of Eq.~(\ref{Exp:1loop}), the
two-point function at one-loop order is summarized as
\begin{align}
 & \langle  \left\{  \gR(X_1),\, \gR(X_2) \right\} \rangle_4 \cr
& \, \Approx \frac{1}{2} \langle \psi^2 \rangle
  \!\int\!  \frac{\dd (\log\! k)}{2\pi^2} \Bigl[\partial^2_{\log k}
 \left\{k^7 \psi_{\bk}(X_1)  \psi^*_{\bk}(X_2)\right\} 
+ \left( {\rm c.c.} \right)  \Bigr] , \cr
\end{align}
where we symmetrized about $X_1$ and $X_2$.
This indicates that all the potentially divergent pieces become the total derivative with
respect to $k$ and hence they vanish
\footnote{Here, we assumed that the ultraviolet
contribution has already been regularized appropriately, say, by 
dimensional regularization. (See Ref.~\cite{Senatore:2009cf}.)
}.

At the leading order in the slow-roll approximation, the condition
(\ref{Exp:SI}), satisfied in the scale-invariant/Bunch Davies vacuum, 
was crucial to remove the IR divergences. 
If we do not choose this vacuum, the quantum state 
is not invariant under the residual gauge transformation, 
and hence the two-point function diverges. 
We think that this possible divergence is attributed to infinitely
large fluctuation in the residual gauge degree of freedom 
corresponding to the overall rescaling of the spatial coordinates. 
This unphysical degree of freedom 
can be tamed if and probably only if we set the initial state to be invariant 
under this residual gauge transformation, as we have anticipated earlier.  

Now, we extend our argument to ${\cal O}(\varepsilon^2)$.  
Once we include the slow-roll corrections, the
condition (\ref{Exp:SI}) no longer ensures the gauge
invariance of the initial state. Using Eqs.~(\ref{Exp:sbR2}),
(\ref{Exp:dx1}), and (\ref{Exp:gR2}), the second-order scalar curvature
is summarized as
\begin{align}
 \gR_2&\Approx \psi \partial^2 \biggl[
    \left(1+\eps_1+\eps_1^2
   + \eps_1 \varepsilon_2 +\lambda_2\right)\partial_\rho\psi
 - (1+\lambda_2) x^i\partial_i\psi  \cr
  &\qquad \quad + \left( \varepsilon_1 + \frac{\varepsilon_2}{2} + 2 \xi_2 \right)\psi
   - \frac{3}{2} {\cal L}^{-1} \varepsilon_2 (2 \varepsilon_1
   + \varepsilon_3) \psi \biggr].  
\end{align}
Here we used 
\begin{eqnarray}
 \partial^2 \delta \zeta_{n,2} \Approx - \frac{3}{2} \left[  \psi
			    \partial^2 {\cal L}^{-1} \varepsilon_2
			    (2\varepsilon_1+\varepsilon_3) \psi  \right],
\end{eqnarray}
which follows from 
\begin{eqnarray*}
 {\cal L} \partial^2 \delta \zeta_{n,2}
  \Approx - \frac{3}{2} {\cal L} \left[  \psi
			    \partial^2 {\cal L}^{-1} \varepsilon_2
			    (2\varepsilon_1+\varepsilon_3) \psi
				 \right]~. 
\end{eqnarray*}

As a natural extension of the condition (\ref{Exp:SI}), we assume that
there is a set of mode
functions which satisfies 
\begin{align}
 &  \left(1+\eps_1+\eps_1^2
   + \eps_1 \varepsilon_2 \right)\partial_\rho v_k
   + \left( \varepsilon_1 + \frac{\varepsilon_2}{2} + 2 \xi_2 \right) v_k \cr
 & \qquad
   - \frac{3}{2}{\cal L}_k^{-1} \varepsilon_2 (2 \varepsilon_1 +
   \varepsilon_3) v_k
  =-D_k v_k+{\cal O}(\varepsilon^3)\,,
\label{Cond:vk}
\end{align}
where we replaced ${\cal L}$ with ${\cal L}_k$:
\begin{equation}
{\cal L}_k :=  \partial_\rho^2 +  (3 - \varepsilon_1+ \varepsilon_2) \partial_\rho
 + \frac{1}{{\rho'}^2}k^2.
\end{equation}
We will show the presence of such mode function in the succeeding
subsection. 

For such mode functions, the 
second-order scalar curvature $\gR_2$ is then simply rewritten as 
$$\gR_2 \Approx - (1+\lambda_2) \psi \partial^2 D_k \psi$$ as in the leading-order of the slow-roll
approximation. Using Eqs.~(\ref{Exp:sbR3}),
(\ref{Exp:dx2}), and (\ref{Exp:gR3}) together with Eq.~(\ref{Cond:vk}), the straight-forward but lengthy
calculation leads to $\gR_3$ in a rather simple expression:
\begin{align}
 \gR_3&\Approx{1\over 2}\psi^2 \partial^2 \biggl[
    (1+ 2 \lambda_2)D_k^2\psi -\mu D_k\psi \cr & \qquad \qquad \qquad +
   \left(- 2 \varepsilon_1^2 + \tfrac{3}{2} \varepsilon_1 \varepsilon_2
 + \tfrac{1}{2} \varepsilon_2 \varepsilon_3 + 6\xi_3\right)\psi
   \biggr] \nonumber\\ &\qquad- \delta \zeta_{n,2} \partial^2 D_k \psi ~.
\end{align}
where we defined 
$\mu:=\varepsilon_1 + \frac{1}{2} \varepsilon_2-3\varepsilon_1^2+ \frac{1}{2} \varepsilon_1
 \varepsilon_2 + 2(\xi_2+\lambda_3) $.
To ensure the absence of the IR divergences, the arbitrary
time-dependent function $\xi_3$ should be chosen as
\begin{eqnarray}
 \xi_3 := \frac{1}{3} \varepsilon_1^2 - \frac{1}{12} \varepsilon_2
  (3\varepsilon_1 + \varepsilon_3)~, \label{Def:xi3}
\end{eqnarray}
to find 
\begin{align}
 ^g\! R_3& \Approx {1\over 2}\psi^2 \partial^2 \left[ (1+ 2\lambda_2)
    D_k^2\psi -\mu D_k\psi  \right] - \delta \zeta_{n,2} \partial^2 D_k \psi ~.
\end{align}
The possibly divergent terms are then summarized as 
\begin{align}
 &  \langle \left\{  {}^g\!R(X_1),\, {}^g\!R(X_2) \right\} \rangle_4 \cr
 &~ \Approx \frac{1}{2} \langle \psi^2 \rangle
\!\int\! {\dd(\log k) \over 2 \pi^2} \bigl\{ 
   (1+ 2 \lambda_2) \partial_{\log k}^2 
   - \mu  \partial_{\log k} 
   \bigr\}  \cr & \qquad \qquad \qquad \qquad \qquad  \times\!  \left\{ \left(k^7 \psi_{\bk}(X_1)
 \psi_{\bk}^*(X_2)\right)\! + \left( {\rm c.c.} \right) \right\}
   \cr
 &  \quad - \langle \delta \zeta_{n,2} \rangle \!\int\!  {\dd(\log k) \over 2 \pi^2}
 \partial_{\log k}\! \left\{ \left(k^7 \psi_{\bk}(X_1)
 \psi_{\bk}^*(X_2)\right)\! + \left( {\rm c.c.} \right) \right\},\cr 
\end{align}
indicating that they are completely cancelled. The conditions on the
initial vacuum state are derived by requesting the regularity of the IR
corrections. Since the IR divergence is, in single field models of
inflation, originating from the residual gauge degrees of
freedom~\cite{IRsingle, IRgauge_L}, the regularity conditions can be
considered as the necessary condition for the gauge invariance.

\subsection{Gauge-invariant initial vacuum}
The requirement of the gauge invariance in the initial vacuum leads to
the condition (\ref{Cond:vk}) on the mode function $v_k$. Taking the
mode function $v_k$ in a similar form to Eq.~(\ref{Exp:vk_BD}) as
\begin{eqnarray}
 v_k(\bar\rho) = \frac{{\rho'}^2 e^{-\rho}}{\phi'} \frac{1}{k^{3/2}} f_k(\bar\rho)~, 
\end{eqnarray}
the condition (\ref{Cond:vk}) can be recast into a rather simple form:
\begin{eqnarray}
 && \left( \partial_{\bar\rho} + \partial_{\log k} \right)
  f_k(\bar\rho) + 
  \left(2 \xi_2-\varepsilon_1^2-{1\over 2}\varepsilon_1\varepsilon_2
   \right) f_k(\bar{\rho}) \cr
 && \qquad \qquad \qquad  - \frac{3}{2} 
   {\cal L}_k^{-1} \varepsilon_2 (2
  \varepsilon_1 + \varepsilon_3) f_k(\bar\rho) 
   = 0~, \label{Cond:fk}
\end{eqnarray} 
where we changed the time variable $\rho$ into 
\begin{eqnarray}
  \bar{\rho} = 
   \log\rho' - \varepsilon_1 + {\cal O}(\varepsilon^2)~.
\end{eqnarray}
Notice that $\bar\rho$ is approximately identical to 
$\rho$ in the sense $\dd\bar\rho/\dd\rho=1+{\cal O}(\varepsilon)$.

The mode equation ${\cal L}_k v_k=0$ yields the evolution equation of
$f_k$ as
\begin{eqnarray}
 \bar{\cal L}_k f_k(\bar\rho) =0~, 
\label{Eq:fk}
\end{eqnarray}
where 
\begin{align}
 \bar{\cal L}_k &= \partial^2_{\bar\rho} + 3 \partial_{\bar\rho}
  + e^{-2(\bar\rho - \log k)} \cr
 & \qquad \qquad  \quad - 3 (\varepsilon_1 + \varepsilon_2/2)
  + {\cal O}\left( \varepsilon^2 \right).
\label{calL}
\end{align}
This operator $\bar{\cal L}_k$ is identical to ${\cal L}_k$ 
at the leading order in the slow roll approximation. 
To fix the form of the terms written as ${\cal O}(\varepsilon^2)$ 
in the above equation, 
we need to specify the form of $\bar\rho$ up to
${\cal O}(\varepsilon^2)$. Since the explicit forms 
of the terms of ${\cal O}(\varepsilon^2)$ in Eq.~(\ref{calL})
are not necessary for the following discussion,  
we leave the higher order corrections to 
$\bar\rho$ unspecified here. 
Operating $\partial_{\bar\rho}+\partial_{\log k}$ on
Eq.~(\ref{Eq:fk}), the solution of the mode equation (\ref{Eq:fk}) is
found to satisfy 
\begin{eqnarray}
 \bar{\cal L}_k \left( \partial_{\bar\rho} + \partial_{\log k}
		\right)  f_k = \frac{3}{2}\varepsilon_2(2\varepsilon_1 +
 \varepsilon_3) f_k~, \label{Cond:fkv2}
\end{eqnarray} 
where we used 
$[\partial_{\bar\rho}+\partial_{\log k},\,\bar{{\cal L}}_k]=-3\varepsilon_2(\varepsilon_1+\varepsilon_3/2)$.
Multiplying the inverse of ${\cal L}_k$ on Eq.~(\ref{Cond:fkv2}),
Eq.~(\ref{Cond:fkv2}) reproduces the 
gauge-invariance condition (\ref{Cond:fk}), where the second term
of Eq.~(\ref{Cond:fk}) appears as a homogeneous solution of 
$\bar{\cal L}_k$.
This indicates that the solution of the mode equation
(\ref{Eq:fk}) consistently satisfies the gauge-invariance condition.

It is also possible to show that the gauge-invariance condition 
(\ref{Cond:fk}) is 
sufficient to ensure that the mode equation is 
satisfied for all wavenumbers, if it is satisfied 
for a particular wavenumber $k_0$:
${\cal L}_{k_0} f_{k_0}(\bar\rho)=0$. 
In fact, by using the gauge-invariance
condition, one can show 
\begin{eqnarray}
&&\partial_{\log k}   \bar{\cal L}_{k} f_{k}|_{k=k_0}
 =\left( \partial_{\bar\rho} + \partial_{\log k} \right)
  \bar{\cal L}_{k} f_{k}|_{k=k_0} 
\nonumber\\ &&\, = 
 \bar{\cal L}_{k} 
 \left( \partial_{\bar\rho} + \partial_{\log k} \right)
  f_{k}|_{k=k_0}-{3\varepsilon_2\over
  2}\left(2\varepsilon_1+\varepsilon_3 \right)f_{k_0} 
 +{\cal O}(\varepsilon^3) \nonumber\\
&&\, =  {\cal O}(\varepsilon^3)~, 
\end{eqnarray}
which proves that thanks to the gauge-invariance condition, the mode function for the another wavenumber
is guaranteed from that for $k_0$.

As was anticipated in Sec.~\ref{SSec:Proof}, 
the commutation relation for $\psi$ is now verified. 
This commutation relation is equivalent 
to the normalization condition for $v_k$ given by 
\begin{eqnarray*}
 {\cal N}_{k} :=  
{i e^{3\bar\rho} 
\left(
f_{k} \partial_{\bar\rho}f_{k}^* - f_{k}^*
  \partial_{\bar\rho}f_{k} \right)
\over
    k^3 
 \left\{ 1 -2\varepsilon_1 +{\cal O}(\varepsilon^2)\right\}}  
= 1~. 
\end{eqnarray*}
As in the case of mode equation, we assume that the normalization condition is
satisfied for a particular wavelength $k_0$ as
$ {\cal N}_{k_0}=1$. 
Then, using the gauge-invariance condition (\ref{Cond:fk}), we obtain
\begin{eqnarray}
\partial_{\log k} {\cal N}_{k}|_{k=k_0}&=& 
 \left( \partial_{\bar\rho} + \partial_{\log k} \right)  {\cal
 N}_{k}|_{k=k_0} 
\nonumber\\ &=& B - 4 \xi_2~.
\label{eq90}
\end{eqnarray}
where the first term $B$ is of ${\cal O}(\varepsilon^2)$ and its
explicit form is not necessary in the current discussion.  
An important fact is 
\begin{eqnarray}
\partial_{\bar\rho}\partial_{\log k} {\cal N}_{k}|_{k=k_0}=0,  
\end{eqnarray}
holds exactly, because $\partial_{\bar\rho}$ and 
$\partial_{\log k}$ commute with each other and the normalization
condition ${\cal N}_k$ is conserved. Therefore the right hand side of 
Eq.~(\ref{eq90}) is guaranteed to be constant in time. 
Then, by choosing $\xi_2$ appropriately, we can always set 
\begin{eqnarray}
\partial_{\log k} {\cal N}_{k}|_{k=k_0}&=& 
{\cal O}\left(\varepsilon^3 \right)~. 
\end{eqnarray}
This proves that one can extend
the mode function by the 
gauge-invariance condition (\ref{Cond:fk}) to the other wavenumbers
keeping the normalization condition satisfied.

We summarize how the time dependent functions 
$\xi_2$, $\xi_3$, $\lambda_2$ and $\lambda_3$ are determined 
uniquely and consistently.  
$\xi_2$ was fixed by requesting 
the normalization condition to be consistent with 
the gauge invariance of the initial state, while 
$\xi_3$ was fixed from the IR regularity of 
the two point function. 
As presented in Eq.~(\ref{commutator}), to ensure the
consistent commutation relations, $\lambda_2$ and $\lambda_3$ 
are also fixed once $\xi_2$ and $\xi_3$ are given. 
In this paper, 
we have not derived the gauge-invariance condition (\ref{Cond:fk}) 
but we just postulated it.  
The above discussions, however, have proven that this condition can 
be imposed consistently by
choosing the homogeneous solution appropriately in $\zeta_n$.

\section{Conclusion}
  \label{Sec:Conclusion}

We presented, in the standard single field inflation model, one example of the calculation of a 
genuine gauge-invariant quantity, i.e., 
the two-point function of $\gR$, which is the spatial 
curvature perturbation on a $\phi$=constant hypersurface 
with its arguments specified in terms of the geodesic 
normal coordinates. 
We showed that, taking an appropriate initial vacuum, 
the two-point function for $\gR$ no
longer yields IR divergences at one-loop order. It would be also
possible to extend our argument to higher orders in loops and also to the
general $n$-point functions. 
The quantities that are compared with actual observations 
like the fluctuation in the Cosmic Microwave
Background should also be such genuine gauge-invariant quantities. 
Hence, our result strongly indicates that such quantities are 
also IR regular for the standard single field inflation model.

In the global gauge that we used in this paper 
the residual gauge degrees of freedom were not fixed. 
The residual gauge degrees of freedom include the overall 
spatial scale transformation corresponding to a constant shift of 
$\zeta$ in the $\delta\phi=0$ gauge, which is the origin 
of the IR divergences. To remove IR divergences, hence, 
we had to impose the invariance of quantum states 
in the direction of the residual gauge, which requests additional 
gauge invariance conditions, such as Eqs.~(\ref{Cond:vk}) and (\ref{Def:xi3}),
on the choice of the initial quantum state. 
The condition (\ref{Cond:vk})
restricts the mode function for the interaction picture field and the condition
(\ref{Def:xi3}) restricts the relation between the Heisenberg field and
the interaction picture field.
In the present paper, we derived these conditions, requiring, instead of
the gauge-invariance itself, that the possibly divergent terms should be
written in the form of total derivatives under the momentum
integral. While the adopted iterative solution to the Heisenberg
equation, that is used to obtain these gauge invariance conditions, may
not be general enough. In this sense, it may be possible to find other
vacua that are regular against IR contributions. Namely, the
gauge-invariance condition (=regularity condition) may not uniquely
determine the vacuum state in the inflationary universe.
Therefore what we have derived is not a necessary
condition but a sufficient condition for the IR regularity. 
We also expect one can specify the initial quantum 
state directly from requirement of the gauge-invariance,  
but this point has not been clarified yet at all. 
We leave these issues for future work. 
At the leading order in the slow-roll approximation, the condition
(\ref{Cond:vk}) is automatically satisfied in the scale-invariant
(Bunch-Davies) vacuum. However, at the higher order in the 
slow-roll approximation,
this condition looks quite non-trivial. 
Our preliminary analysis tells that 
this vacuum state seems to coincide with 
the one naturally obtained by using the usual 
$i\epsilon$ prescription. 
We would like to report on this point in our 
forthcoming publication~\cite{IR_vacuum}.

In contrast to the case of the global gauge, if we fix all the residual gauge degrees of freedom and quantize only physical degrees of freedom, we need not to restrict the initial quantum state by imposing additional gauge invariance conditions. This would provide another way of quantization that also yields no artificial divergences. In our previous work~\cite{IRsingle}, following this direction, we tried to fix the residual gauge degrees of freedom, including the overall scaling of spatial coordinates, by imposing appropriate conditions. (We refer to this gauge as the local gauge~\cite{IRsingle}.) If we perform the canonical quantization in the local gauge and give the initial state there, the local gauge conditions ensure the regularity of the IR corrections, without restricting the initial quantum state. It is, however, not so trivial to perform canonical quantization in the local gauge, because of additional conditions to fix the residual gauge degrees of freedom. Furthermore, if we set initial vacuum in the local gauge, it would be in general breaks the invariance under spatial translation. In our previous work \cite{IRsingle}, 
We therefore chose the initial vacuum state in the global gauge and then linearly
transformed the interaction picture field from the global gauge to the local one.
The truth is that in this gauge transformation there appear the non-linear
contributions that can cause IR divergence. These contributions were not
taken into account in Ref.~\cite{IRsingle} and the expression in the
local gauge was still including the divergent terms. One can say that
this is due to the lack of the gauge invariance in choosing the initial
state, because the IR divergence actually disappears at the leading
order in the slow-roll approximation if we take the Bunch-Davies vacuum,
which is invariant under the scale transformation. To guarantee the IR
regularity also at higher orders in this approximation, we need to adapt
the gauge-invariance condition as is done in this paper~\footnote{These
points will be clarified in the errata of \cite{IRsingle}.}.

In this paper we have solved the Heisenberg equation 
to the second order in the slow roll approximation. 
Up to this order, we showed the presence of a gauge invariant 
initial quantum state that is free from IR divergences.  
But it looks quite non-trivial to extend our results 
to the higher order in the slow roll approximation. 
For the complete understanding of IR issue 
in the standard single field inflation model, 
it is also definitely necessary to prove 
the existence of such an initial quantum state 
without relying on the slow roll approximation.   

\acknowledgments
The discussions during the workshops YITP-T-09-05, YITP-T-10-01, and YKIS2010 at Yukawa Institute were
very useful to complete this work. Y.~U. and T.~T. would like
to thank Arthur Hebecker for his valuable comments. Y.~U. and
T.~T. are supported by
the JSPS under Contact Nos.\ 22840043 and 21244033. We also acknowledge the support
of the Grant-in-Aid for the Global COE Program ``The Next Generation of
Physics, Spun from Universality and Emergence'' and the Grant-in-Aid for
Scientific Research on Innovative Area Nos.\ 21111006 and 22111507 from the MEXT. 

\appendix
\begin{widetext}
\section{Third-order gauge transformation} 
\label{Sec:transformation}
In this appendix, we study the change of variables between the flat
gauge with $\sR=0$ and the comoving
gauge with $\delta \phi=0$. We denote the 
gauge transformation from the flat gauge to the comoving gauge as
$\tilde{x}^\mu = e^{{\cal L}_\xi} x^\mu$ where $\xi^\mu$ is
the vector field $\xi^\mu = (\alpha, \beta^i)$. 
In the present paper 
it is not strictly necessary to transform the spatial coordinates, 
but we do this for future application. 
For the same reason, we also leave the terms
that do not yield IR divergences here. 
 
In accordance with
Ref.~\cite{Maldacena2002}, we use the cosmic time $t$
and denote its derivative by a dot. From 
the condition that this transformation
makes the scalar field perturbation to vanish, we obtain 
\begin{eqnarray}
 0 = \varphi + {\cal L}_\xi (\phi+ \varphi)+
 \frac{1}{2!} {\cal L}_{\xi}^2 (\phi+ \varphi) + 
 \frac{1}{3!} {\cal L}_{\xi}^3 (\phi+ \varphi) + \cdots~,
 \label{Exp:Transphi}
\end{eqnarray}
It is enough to calculate the perturbed expansion up to the third order.
Using Eq.~(\ref{Exp:Transphi}), the time shift
$\alpha$ is solved at each order as 
\begin{eqnarray}
 \dot{\rho} \alpha_1 &=& - \frac{\dot{\rho}}{\dot{\phi}} \varphi =: 
  \zeta_n~, \label{Exp:alpha1} \\
  \dot{\rho} \alpha_2 &=& \frac{\dot{\rho}}{2 {\dot\phi}^2} \varphi \dot{\varphi}=
  \frac{1}{4} ( \partial_\rho + \varepsilon_2) \zeta_n^2~,  \label{Exp:alpha2} \\
  \dot{\rho} \alpha_3 
  &=&  \frac{1}{12} \zeta_n^2 \partial^2_\rho \zeta_n 
 + \frac{1}{3} \zeta_n \left( \partial_\rho \zeta_n \right)^2
 + \frac{3}{16} \varepsilon_2 \zeta_n \partial_\rho \zeta_n^2
 + \frac{1}{24}
  \varepsilon_2(2\varepsilon_2+\varepsilon_3) \zeta_n^3
 + \frac{1}{2} \beta^i_2 \partial_i \zeta_n ~,  \label{Exp:alpha3}
\end{eqnarray}
where, $\beta_1=0$ is presumed. 
In both flat and comoving gauges, the traceless
part of the spatial metric 
$g_{ij}$ is requested to satisfy the transverse conditions. To
maintain these transverse conditions, 
we also need to change the spatial coordinates at second 
order. The spatial component of the metric then transforms as 
\begin{eqnarray}
 && e^{2\zeta} \left[e^{\delta \gamma}\right]_{ij} = \left[ e^{\delta
		\tilde{\gamma}} \right]_{ij} 
 + (\alpha_1+ \alpha_2+\alpha_3)
  e^{-2\rho} \partial_t \left(e^{2\rho}\left[e^{\delta \tilde{\gamma}}\right]_{ij} \right) 
 + \beta^k_2 \partial_k \left[ e^{\delta \tilde{\gamma}} \right]_{ij}
 + 2 \left\{ \partial_{(i} \beta^k_2 + \partial_{(i} \beta^k_3 \right\} \left[
    e^{\delta \tilde{\gamma}} \right]_{j)k} \cr
 && \qquad \qquad \qquad + \frac{1}{2} e^{-2\rho} (\alpha_1+\alpha_2) \partial_t
 \left\{ (\alpha_1+\alpha_2) \partial_t \left( e^{2\rho} \left[
							  e^{\delta
							  \tilde{\gamma}}
							 \right]_{ij}
					\right) \right\}
 + \frac{1}{3} e^{-2\rho} \alpha_1 \partial_t \left\{ \alpha_1
					       \partial_t (\alpha_1
					       \dot{\rho} e^{2\rho})
					      \right\} \delta_{ij}
 + {\cal H}_{ij} , \label{Exp:Transhij2}
\end{eqnarray}
where ${\cal H}_{ij}$ is defined as
\begin{eqnarray}
 {\cal H}_{ij} &:=& 2e^{-2\rho} \left\{ \partial_{(i} \alpha_1 + \partial_{(i}
		     \alpha_2  \right\}  N_{j)}
 +  \alpha_1 e^{-2\rho} \partial_t \left\{ \partial_{(i} \alpha_1 N_{j)} 
 + e^{2\rho} \partial_{(i} \beta_{j),2}\right\}
 +  \beta^k_2 \partial_k \dot{\rho} \alpha_1 \delta_{ij}
 + 2 \dot{\rho} \alpha_1 \partial_{(i} \beta_{j),2} \cr
 && \qquad \quad + e^{-2\rho} \partial_{(i} \alpha_1 
  \left\{ - \partial_{j)} \alpha_1 - 2 \partial_{j)} \alpha_2 - 2
   \partial_{j)} \alpha_1 \delta N + 
 \alpha_1 \dot{N}_{j)} + \dot{\alpha}_1 N_{j)} +  e^{2\rho}
 \dot{\beta}_{j),2} \right\} \cr
 && \qquad \quad - \alpha_1 \partial_{(i} \alpha_1 \partial_{j)}
 \dot\alpha_1 - \dot{\alpha}_1 \partial_i \alpha_1 \partial_j \alpha_1~.
\end{eqnarray}
Taking the trace part of Eq.~(\ref{Exp:Transhij2}), we obtain
\begin{eqnarray}
 \zeta &=& \dot{\rho} \left( \alpha_1 + \alpha_2+\alpha_3 \right)
 + \left\{ \dot{\rho} (\alpha_1+\alpha_2) \right\}^2
 + \frac{1}{2} (\alpha_1+\alpha_2) \partial_t \left\{ \dot{\rho}(\alpha_1+\alpha_2)
				   \right\}
 + \frac{1}{6} e^{-2\rho} \alpha_1 \partial_t \left\{ \alpha_1
					       \partial_t (\dot{\rho}
					       \alpha_1 e^{2\rho})
					      \right\} \cr
 && \quad  - (\dot{\rho}\alpha_1)^2 + \frac{4}{3} (\dot{\rho}
 \alpha_1)^3 - \frac{1}{6} \delta^{ij} 
 \left\{ \left[ e^{\delta \gamma} \right]_{ij} - 
   \left[ e^{\delta \tilde\gamma} \right]_{ij} \right\} 
 + \frac{1}{6} \alpha_1 \partial_t\, \delta^{ij}\! \left[ e^{\delta
						\tilde\gamma}
					       \right]_{ij}
 + \frac{1}{6} {\cal H} \cr
 && \quad + \frac{1}{3} (\partial^i \beta^j_2 + \partial^i \beta^j_3)
 \left[ e^{\delta \tilde{\gamma}} \right]_{ij}
  - \dot{\rho} \alpha_1 \left\{ 2 \dot{\rho} \alpha_2 
 + \frac{2}{3} \partial^i \beta_{i,2} +
 \alpha_1 e^{-2\rho} \partial_t (\dot{\rho} \alpha_1 e^{2\rho}) +
 \frac{1}{3} {\cal H}  \right\}~, \label{Exp:zeta}
\end{eqnarray}
where we defined ${\cal H} := \delta^{ij} {\cal H}_{ij}$. For our
purpose, it is sufficient to consider the gravitational wave 
perturbation up to the
second order. Neglecting the third-order terms, the transformation of
the transverse traceless tensor is given by
\begin{eqnarray}
 \delta \gamma_{ij} 
 &=& \delta \tilde{\gamma}_{ij} + \zeta_n \partial_\rho \delta
 \tilde{\gamma}_{ij} +\left( {\delta_i}^k {\delta_j}^l +  {\delta_j}^k {\delta_i}^l -
    \frac{2}{3} \delta_{ij} \delta^{kl} \right)
  \left( \partial_k \zeta_n \frac{e^{-2\rho}}{\dot{\rho}} N_l + \partial_k \beta_{l,2} 
 - \frac{1}{2} \frac{e^{-2\rho} }{\dot{\rho}^2} \partial_k \zeta_n
 \partial_l \zeta_n   \right) ~. \label{Exp:Transgamma}
\end{eqnarray}
Now, it is clear that $\delta \gamma_{ij}$ agrees with 
$\delta \tilde{\gamma}_{ij}$, after we neglect the terms that are irrelevant
to IR divergences. 
Using
Eqs.~(\ref{Exp:alpha1})-(\ref{Exp:alpha3}) and Eq.~(\ref{Exp:Transgamma}), Eq.~(\ref{Exp:zeta}) is
rewritten as 
\begin{eqnarray}
 \zeta &=& \zeta_n + \frac{1}{2} \partial_\rho \zeta_n^2 +
  \frac{1}{4} \varepsilon_2 \zeta_n^2 + \frac{1}{2} \zeta_n^2
  \partial^2_\rho \zeta_n + \frac{3}{8} \varepsilon_2 \zeta_n
  \partial_\rho \zeta_n^2 + \frac{1}{12} \varepsilon_2 (\varepsilon_2 +
  2\varepsilon_3) \zeta_n^3+
  \zeta_n (\partial_\rho
  \zeta_n)^2 + \frac{1}{2} \beta_2^i \partial_i \zeta_n -
  \frac{2}{3}\zeta_n \partial^i \beta_{i,2} \nonumber \\
 && \qquad + \frac{1}{3} (\partial_i \beta^i_2 + \partial_i \beta^i_3)
 - \frac{1}{3} \delta \tilde{\gamma}^{ij} \partial_i \zeta_n  \frac{e^{-2\rho} }{\dot{\rho}^2} N_j
  + \frac{1}{6} {e^{-2\rho} \over \dot{\rho}^2 } \delta
  \tilde{\gamma}^{ij} \partial_i \zeta_n \partial_j \zeta_n
 +  \frac{1}{6} (1-2\zeta_n) {\cal H}\,.
\end{eqnarray}
Multiplying
the spatial derivative $\partial^i$ on Eq.~(\ref{Exp:Transhij2}), $\beta_{i,2}$ is given as a
solution of the Poisson equation:
\begin{eqnarray}
 \partial^2 \beta_{i,2}= - \partial^j \zeta_n \partial_\rho \delta
  \tilde{\gamma}_{ij} - \partial^j {\cal H}_{ij,2} + \frac{1}{3}
  \partial_i {\cal H}_2 + \frac{1}{4} \partial_i \partial^{-2} 
 \left[ \partial^k \partial^l \zeta_n \partial_\rho \delta
  \tilde{\gamma}_{kl} + \left( \partial^k \partial^l - \frac{1}{3}
			 \delta^{kl} \partial^2 \right){\cal H}_{kl,2} \right]~, 
\end{eqnarray}
where we used $\partial^i \beta_{i,2}$, given by operating $\partial^i \partial^j$ on
Eq.~(\ref{Exp:Transhij2}). At the third order, $\beta_{i,3}$ is obtained
in a similar manner. It is notable that $\beta_{i,n}\,(n=2,3,\cdots)$ is
multiplied by at least one wavenumber vector in the momentum
representation. 
Neglecting the terms that are irrelevant to the IR
divergences, 
the curvature perturbation $\zeta$ is related to $\zeta_n$ as
\begin{eqnarray}
 \zeta = \zeta_n + \frac{1}{2} \partial_\rho \zeta_n^2 +
  \frac{\varepsilon_2}{4} \zeta_n^2 + \frac{1}{2} \zeta_n^2
  \partial^2_\rho \zeta_n + \frac{3}{8} \varepsilon_2 \zeta_n
  \partial_\rho \zeta_n^2 + \frac{1}{12} \varepsilon_2 (\varepsilon_2 +
  2\varepsilon_3) \zeta_n^3 + \cdots ~.
\end{eqnarray}
Here, we also neglected the cubic-order terms 
with only one graviton field $\delta \gamma_{ij}$, since its 
contribution vanishes in $\langle \gR \gR \rangle_4$. 

\end{widetext}

\end{document}